\documentclass[twocolumn,prb,aps,notitlepage]{revtex4-1}
\usepackage{tikz}
\usepackage{amssymb}
\usepackage{psfrag}
\usepackage{textcomp}
\usepackage{pdfsync}
\usepackage{amsmath}
\usepackage{amsfonts}
\usepackage{graphicx,bm,epstopdf}
\usepackage{subfigure}
\usepackage{comment}
\usepackage{verbatim}
\usepackage{color}
\usepackage{upgreek}
\usepackage{tikz}
\usepackage{natbib,hyperref}
\usepackage{ifpdf}
\usepackage{float}
\usepackage[normalem]{ulem}

\begin{document}

\title{Chiral surface and edge plasmons in ferromagnetic conductors}
\author{Steven S.-L. Zhang$^{1,2}$}
\email{shulei.zhang@anl.gov}
\author{Giovanni Vignale$^1$}
\email{vignaleg@missouri.edu}

\affiliation{$^1$Department of physics and astronomy, University of
Missouri, Columbia, Missouri, 65211, USA\\$^2$Materials Science Division,
Argonne National Laboratory, Lemont, Illinois, 60439, USA}
\date{\today }

\begin{abstract}
The recently introduced concept of ``surface Berry plasmons" is studied in
the concrete instance of a ferromagnetic conductor in which the Berry
curvature, generated by spin-orbit (SO) interaction, has opposite
signs for carriers parallel or antiparallel to the magnetization. By using
collisionless hydrodynamic equations with appropriate boundary conditions,
we study both the surface plasmons of a three-dimensional ferromagnetic
conductor and the edge plasmons of a two-dimensional one. The anomalous
velocity and the broken inversion symmetry at the surface or the edge of the
conductor create a ``handedness", whereby the plasmon frequency depends not
only on the angle between the wave vector and the magnetization, but also on
the direction of propagation along a given line. In particular, we find that
the frequency of the edge plasmon depends on the direction of propagation
along the edge. These Berry curvature effects are compared and contrasted
with similar effects induced in the plasmon dispersion by an external
magnetic field in the absence of Berry curvature. We argue that Berry
curvature effects may be used to control the direction of propagation of the
surface plasmons via coupling with the magnetization of ferromagnetic conductors, and thus create a link between plasmonics and spintronics.
\end{abstract}

\maketitle

\section{Introduction}

The discovery of collective oscillations of electrons in quantum solid-state
plasmas in the 1950s was a major milestone in the evolution of condensed
matter physics~\cite{noz_pin_tfl}. It exposed the fundamental dichotomy in
the character of electronic elementary excitations, which can be either
individual quasiparticles or organized collective oscillations (plasmons),
and spawned a variety of theoretical treatments of the electron gas (the RPA~%
\cite{noz_pin_tfl,qtel} being one of the earliest and most successful),
effectively igniting the field of many-electron physics~\cite{Mahan}. By the
end of the 20th century the interest began to shift to the possible
technological applications of plasmons, as it was realized that the
wavelengths of these oscillations, being much shorter then the wavelength of
light at the same frequency, could be used to compress electromagnetic
energy to a nanometric scale -- the scale of integrated circuits and
devices. A thriving area of research, known as ``plasmonics"~\cite%
{Maier03,Stockman11}, was born.

At about the time that plasmonics was taking off, major advances were made
in the band theory of solids~\cite{AM76,LL80}. It was realized that, under
quite common conditions, the Bloch wave functions of electrons in a periodic
solid, regarded as functions of the Bloch wave vector $\mathbf{k}$ in the
Brillouin zone, have non-trivial geometric properties. When the $n$-th
eigenstate $|u_{n}(\mathbf{k})\rangle $ of the periodic Hamiltonian $H(%
\mathbf{k})$ is adiabatically transported around a closed loop in the
Brillouin zone the final state differs from the initial one by a gauge
invariant \textquotedblleft Berry phase" $\Delta \phi $, which equals the
flux of \textquotedblleft Berry curvature" through the area enclosed by the
loop~\cite{Berry84}. The mathematical expression for the Berry curvature
\begin{equation}
\boldsymbol{\Omega }_{n}\left( \mathbf{k}\right) =i\left\langle \nabla _{%
\mathbf{k}}u_{n}\left( \mathbf{k}\right) \right\vert \times \left\vert
\nabla _{\mathbf{k}}u_{n}\left( \mathbf{k}\right) \right\rangle
\end{equation}%
is one of the most important properties of a solid-state system, its
integral over the Brillouin zone being connected to topological quantum
numbers and quantized conductivities~\cite{Xiao2010}.

A question that naturally arises at this point is: how does the Berry
curvature of a band affect, if at all, the properties of the plasmons of the
carriers in that band? One of the simplest ways to address the question is
to set up the collisioness hydrodynamic equations for the collective motion
of the electron fluid~\cite{qtel}. These will in turn be based on the
quasiclassical equations of motion for wave packets in the band:~\cite%
{Sundaram99,Xiao2010}
\begin{subequations}
\label{Eq:semi-classical}
\begin{equation}  \label{Eq:semi-classical1}
\mathbf{\dot{r}}=\hbar ^{-1}\nabla _{\mathbf{k}}E_{n}\left( \mathbf{k}%
\right) +\boldsymbol{\Omega }_{n}\left( \mathbf{k}\right) \times \mathbf{%
\dot{k}}
\end{equation}%
\begin{equation}  \label{Eq:semi-classical2}
\mathbf{\dot{k}=-}\hbar ^{-1}\nabla _{\mathbf{r}}V\left( \mathbf{r}\right)\,,
\end{equation}%
where $\mathbf{r}$ and $\hbar $\thinspace $\mathbf{k}$ are the position and
the momentum of the wave packets, $E_{n}\left( \mathbf{k}\right) $ is the
energy of the Bloch state, and $V\left( \mathbf{r}\right) $ is the potential
energy arising from the self-consistent electric field. The Berry curvature
enters the equations of motion through the second term in the expression for
$\mathbf{\dot{r}}$. This is often referred to as the ``anomalous velocity":

\end{subequations}
\begin{equation}
\mathbf{v}_{a}\left( \mathbf{k}\right) \equiv \boldsymbol{\Omega }_{n}\left(
\mathbf{k}\right) \times \mathbf{\dot{k}=-}\hbar ^{-1}\boldsymbol{\Omega }%
_{n}\left( \mathbf{k}\right) \times \nabla _{\mathbf{r}}V\left( \mathbf{r}%
\right)\,.
\end{equation}

Physically, the anomalous velocity reflects the non-conservation of the
Bloch momentum. As the Bloch momentum changes under the action of a force
according to Eq.~(\ref{Eq:semi-classical2}), the quantum state of the
electron no longer coincides with the instantaneous eigenstate $\left\vert
u_{n}\left( \mathbf{k}\left( t\right) \right) \right\rangle $. \ The
difference between the actual state $\left\vert u_{n}\left( t\right)
\right\rangle $ and the instantaneous eigenstate $\left\vert u_{n}\left(
\mathbf{k}\left( t\right) \right) \right\rangle $ is reflected in the
expectation value of velocity operator $\mathbf{\hat{v}}\left( \mathbf{k}%
\right) \equiv \hbar ^{-1}\nabla _{\mathbf{k}}\hat{H}\left( \mathbf{k}%
\right) $, where $\hat{H}\left( \mathbf{k}\right) $ is the Hamiltonian at
wave vector $\mathbf{k}$. It is easily seen that the expectation value of $%
\mathbf{\hat{v}}\left( \mathbf{k}\right) $ in the instantaneous eigenstate
equals $\hbar ^{-1}\nabla _{\mathbf{k}}E_{n}\left( \mathbf{k}\right) $. The
anomalous velocity is the correction to that result, arising from the fact
that, in a dynamical situation, $\left\vert u_{n}\left(
t\right)\right\rangle \neq \left\vert u_{n}\left( \mathbf{k}\left( t\right)
\right) \right\rangle $.

The study of the effect of the anomalous velocity on the dynamics of the
plasmons has been pioneered in a recent paper by Song and Rudner (SR)~\cite%
{Song2016}. Working within the framework of collisionless hydrodynamics (see
Section II) they first showed that the anomalous velocity has no effect on
the bulk modes of a homogeneous electron liquid. This is because the
anomalous velocity enters the bulk hydrodynamics only in the continuity
equation, and only through its divergence, which is zero due to $\nabla
_{r}\cdot \left[ \boldsymbol{\Omega }_{n}\left( \mathbf{k}\right) \times
\nabla _{\mathbf{r}}V(\mathbf{r})\right] =0$.

The situation changes when one considers surface or edge plasmons~\cite%
{Ferrell60}. These collective oscillations are exponentially localized near
the surface or the edge of the system, with a localization length of the
order of $v_{F}/\omega _{p}$, where $v_{F}$ is the Fermi velocity and $%
\omega _{p}$ is the plasmon frequency. These are also the modes that are of
greatest interest in plasmonic applications, because they hybridize with
electromagnetic waves to produces surface plasmon polaritons~\cite%
{Barnes2003,Hutter2004,Ebbesen08,Pitarke07}. In the hydrodynamic approach,
surface plasmons are derived by imposing a boundary condition on the current
density at the surface or edge of the system. The boundary condition states
that there is no electron flux through the boundary of the system,
\begin{equation*}
j_{z}(z=0)=0 \,,
\end{equation*}
where $z$ is the direction perpendicular to the boundary and $j_{z}$ is the $%
z$-component of the particle current. Notice that the imposition of the
boundary condition is the way hydrodynamics - a long wavelength theory -
handles the sharp variation of the electronic density across the boundary.
It is precisely through the boundary condition that the anomalous velocity
enters the solution for the plasmon. This point was clearly demonstrated by
SR~\cite{Song2016} for the edge plasmon of a 2D system. Taking a rather
abstract approach in which a Berry curvature of unspecified origin was
assumed to exist, SR showed that the frequency of right-propagating modes
(along the edge) can be significantly different from that of
left-propagating modes. At finite wave vector one of the two modes can be
well defined while its time-reversed partner may be severely Landau-damped.
Under this scenario an essentially unidirectional propagation of edge
plasmons is achieved, which is of great technological interest. The scenario
is similar, but not identical to that of surface and edge plasmons in a
magnetic field (the so-called ``magnetoplasmons"), which were studied, for
example, in Refs.~\cite{Chiu72,Volkov85,Fetter85,Ashoori92}. Both scenarios
require broken time-reversal symmetry to produce chiral plasmons, but the
magnetoplasmon arises from the classical Lorenz force exerted by the
magnetic field, whereas the Berry plasmon arises from the anomalous
velocity. 

In this paper we study a concrete realization of the abstract SR scenario,
namely the Berry plasmons at the surface of a ferromagnetic conductor, with
the magnetization lying in the plane of the boundary surface in 3D or
perpendicular to the plane of the system in 2D. Spontaneous magnetization
breaks time-reversal symmetry, but we assume that the magnetic field
associated with the magnetization has negligible effect on the electrons: in
particular, there is no sizable Lorenz force. On the other hand, the Berry
curvature of electrons in (say) the conduction band is assumed to be
different from zero and spin-dependent, having opposite signs for electrons
of opposite spins. For example, in the conduction band of 3D GaAs, a simple
calculation based on the $8$-band model~\cite{Winkler03} predicts the Berry
curvature
\begin{equation}
\boldsymbol{\Omega }_{c}(\mathbf{k})\simeq \lambda ^{2}\boldsymbol{\sigma }
\end{equation}
at the bottom of the band. The effective Compton wavelength $\lambda $ is
related to band parameters by the well-known formula
\begin{equation}
\lambda ^{2}=\frac{2\hbar ^{2}|P|^{2}}{3m_e^{2}}\left[ \frac{1}{E_{g}^{2}}-%
\frac{1}{(E_{g}+\Delta )^{2}}\right]\,,
\end{equation}%
where $E_{g}$ is the fundamental band gap, $\Delta $ is the gap separating
the light/heavy hole bands from the so-called spin-orbit (SO) split band,
and $P$ is the matrix element of the momentum operator between atomic $s$
and $p$ states. Thus, the $\Delta $ gap is a direct measure of the
SO-induced splitting of atomic energy levels with $J=1/2$ and $3/2$, and its
non-zero value is essential to the emergence of a finite Berry curvature in
the conduction band. Because electrons of opposite spins have opposite Berry
curvatures, and hence opposite anomalous velocities, no effect is expected
on the surface plasmons of a spin-unpolarized system. But, if the system is
magnetic, then the opposite anomalous velocities of majority and minority
spin electrons give a non-vanishing contribution to the net particle
current, which affects the collective motion of the electron liquid, and
maximally so when the electron liquid is fully spin polarized. We refer to
these collective motions as \textit{ferromagnetic surface (or edge) plasmons}%
.

The results of our study, presented below, pertain to long-wavelength
plasmons (wave vector $q \ll k_F$, where $k_F$ is the Fermi wave vector),
but the wavelength is not so large that retardation effects must be taken
into account: namely, we assume $q \gg \omega_p/c$, where $c$ is the speed
of light. Even in this limit, we find that the frequency of the
ferromagnetic surface plasmon depends on the angle between the wave vector
and the magnetization. Similarly, the frequency of the ferromagnetic edge
plasmon depends on the direction of propagation along the edge. The fact
that charge oscillations, such as the plasmons, ``sense" the magnetization
is a consequence of the spin-orbit-induced Berry curvature. It has nothing
to do with the well known anisotropy of plasmons in a magnetic field. The
relation between chiral magnetoplasmons and chiral ferromagnetic plasmons is
reminiscent of the relation between the regular Hall effect and the
anomalous Hall effect, where the former arises from the Lorentz force, while
the latter arises from the concerted action of spin-orbit coupling and
magnetization. Another significant difference between ferromagnetic surface
plasmons and ordinary magnetoplasmons is that we find a single surface mode,
as opposed to two. The two magnetoplasmons arise from the interplay of two
classical forces, the electrostatic force and the Lorentz force, which can
either work together or against each other. We have no Lorentz force, and
therefore find a single ferromagnetic plasmon.

Chiral edge plasmons have recently been predicted~\cite{Low2015} in
two-dimensional gapped Dirac systems under pumping with circularly polarized
light, which produces a population imbalance between two valleys. By
contrast, the chirality of the plasmons in the present study arises from a
spontaneous spin polarization under equilibrium conditions. A recent study
of topological edge magnetoplasmons~\cite{Jin2016} is not directly relevant
to the present scenario, since it relies on a magnetic field rather than a
spontaneous magnetization.

The angular dependence and chirality of the plasmon frequency is a
potentially important issue in plasmonics, since it can be used to control
the direction of propagation of plasmon waves. Even more interesting, in our
view, is the unusual coupling between magnetism and charge oscillations that
this work foreshadows. The coupling should persist in fully dynamical
situations, when both the magnetization and the charge density are time
dependent. This suggests the intriguing possibility of coupling plasmons and
spin waves, thus bringing together the fields of spintronics~\cite%
{Wolf01,Zutic04,Fabian07,Wu2010,np_3_153} and plasmonics.

The remaining of the paper is organized as
follows. In Sec.~\ref{sec:formulation}, we introduce a hydrodynamic model of electron fluid to investigate the dispersion of surface (or edge) plamson in ferromagnetic conductors. Our treatment is based on a set of linearized hydrodynamic equations and the corresponding boundary conditions for the surface or edge plasmon modes in which the anomalous velocity comes into play. We then present the exact solution of the 3D ferromagnetic surface plasmon dispersion in Sec.~\ref{sec:3D-sol}. An approximate solution of the 2D problem will be discussed in Sec.~\ref{sec:2D-sol} with a simplified treatment of the electrostatics. Following the thorough investigation of both the ferromagnetic surface and edge plamsons, we show, in Sec.~\ref{Sec:discussion}(A), that they are distinguishable from the classical surface and edge magnetoplasmons (arising from the Lorentz force) by providing a comparison between the two kinds of plasmons in the long wavelength limit. And finally in Sec.~\ref{Sec:discussion}(B), we discuss possible experimental observation of the ferromagnetic surface and edge plasmons in various ferromagnetic systems. The conclusion is given in Sec.~\ref{sec:conclusion}.

\section{Collisionless hydrodynamics}
\label{sec:formulation}

Unlike proper hydrodynamics, which presupposes slow collective motion on the
scale of the particle-particle collision frequency, collisionless
hydrodynamics applies to high-frequency collective motion, such as plasmons,
in which collisions between quasiparticle can be disregarded~\cite%
{noz_pin_tfl}. It is well known that the full-fledged collisionlesss
hydrodynamic treatment must include a viscoelastic stress tensor, which
produces not only the ``hydrostatic force" (gradient of pressure), but also
an elastic shear force and a viscous friction force~\cite{qtel}. These two
additional forces are essential to obtain, respectively, the correct
dispersion of plasmons at finite wave vector and the non-Landau damping~\cite%
{noz_pin_tfl}. In this paper, however, we limit ourselves to a more crude
model, in which we neglect shear and viscous forces. This approach is
expected to become essentially exact in the long wavelength limit, due to
the dominance of electrostatic forces, but will become inaccurate at shorter
wavelength. These inaccuracies are of secondary importance here, since our
primary interest is in the qualitatively new features of the solution, which
appear already in the long wavelength limit. With the above discussion in
mind, the bulk hydrodynamic equations are
\begin{equation}
\partial _{t}\delta n+\nabla \cdot \mathbf{j}=0
\end{equation}%
(continuity equation) and
\begin{equation}
\partial _{t}\mathbf{j}_{p}+s^{2}\nabla \delta n-\frac{en_{0}}{m_{e}}\nabla
\varphi =0
\end{equation}%
the Euler equation, where $n_{0}$ is the uniform equilibrium density of
electrons, $m_{e}$ is the effective mass,
\begin{equation}
\mathbf{j}_{p}=\frac{\mathbf{p}}{m_{e}}
\end{equation}%
is the canonical current density, proportional to the canonical momentum
density $\mathbf{p}$, associated with the momentum variable $\hbar \mathbf{k}
$, $s$ is the velocity of the hydrodynamic sound (this is of the order of
the Fermi velocity, and is related to the bulk modulus $K$ by the well-known
relation $s^{2}=\frac{K}{n_{0}m_{e}}$), $\delta n$ is the deviation of the
electron density from the equilibrium density and $\mathbf{j}$ is the
physical current density given by
\begin{equation}
\mathbf{j}=\mathbf{j}_{p}+\frac{\mathcal{P}n_{0}e}{\hbar }\lambda ^{2}%
\mathbf{\hat{m}\times }\nabla \varphi
\end{equation}%
with $\hat{\mathbf{m}}$ being the unit vector along the magnetization, and $%
\mathcal{P}=\frac{n_{0\uparrow}-n_{0\downarrow}}{n_0}$ the spin polarization
of the electron density.
The hydrodynamic equations contain the electrostatic potential $\varphi $,
which is assumed to be instantaneously created by the charge density
according to the Poisson equation%
\begin{equation}
\nabla ^{2}\varphi =4\pi e\delta n \,.
\end{equation}%
%
%
%
The solution of the Poisson equation is straightforward in 3D, but not at
all in 2D, unless some simplifying approximations are made, which we will
discuss later.

In the absence of boundary conditions it is easy to see that the solution to
the hydrodynamic equations (together with the Poisson equation) is not
affected by the anomalous velocity. This is because, as remarked in the
introduction, $\nabla \cdot (\hat{\mathbf{m}}\times \nabla \varphi )=0$
allows us to replace $\mathbf{j}$ by $\mathbf{j}_p$ in the continuity
equation. Then the coupled equations for $\delta n$ and $\mathbf{p}$ do not
contain the anomalous term and therefore the eigenfrequencies do not depend
on $\lambda $ or $\hat{\mathbf{m}}$. On the other hand, for surface or edge
modes the boundary condition of vanishing charge current density, i.e., $%
\left. \mathbf{j}_{z}\right\vert _{z=0^{-}}=0$, creates a coupling between
charge and magnetization; this boundary condition can be explicitly written as
\begin{equation}
\left[\mathbf{j}_{p,z}+\frac{e\mathcal{P}n_{0}}{\hbar }\lambda ^{2}\left(
\mathbf{\hat{m}\times }\nabla \varphi \right) \cdot \mathbf{\hat{z}}\right]
_{z=0^{-}}=0 \,.  \label{Eq:bc-jz}
\end{equation}
In addition, the electric potential and its gradient must be continuous at $%
z=0$, i.e.,
\begin{equation}
\left. \varphi \right\vert _{z=0^{-}}=\left. \varphi \right\vert _{z=0^{+}}%
\text{ and }\left. \partial _{z}\varphi \right\vert _{z=0^{-}}=\left.
\partial _{z}\varphi \right\vert _{z=0^{+}} \,.  \label{Eq:bc-phi}
\end{equation}

\section{Solution in three dimensions}
\label{sec:3D-sol}

Assuming translational invariance in the plane of the surface, we seek
solutions in the form of plane waves of wave vector $\mathbf{q}$ decaying
exponentially in the bulk ($z<0$) as $e^{\kappa z}$ with $\kappa >0$. We let
all the physical quantities (e.g., $\delta n$, $\mathbf{p}$ and $\varphi $)
take the form $\Psi \left( \mathbf{r},z;t\right) \sim e^{\kappa z}e^{i(%
\mathbf{q}\cdot \mathbf{r}-\omega t)}$ up to some constant coefficients to be determined by the boundary conditions.
Notice that boldface symbols are use to indicate vectors in the plane of the
surface. Inserting this ansatz in the hydrodynamic equations we find a set
of homogeneous algebraic equations, i.e.,
\begin{subequations}
\begin{equation}
-i\omega \delta n+\kappa j_{p,z}+i\mathbf{q}\cdot \mathbf{j}_{p,\parallel }=0
\label{Eq:H1}
\end{equation}%
\begin{equation}
-i\omega j_{p,z}+s^{2}\kappa \delta n-\frac{en_{0}}{m_{e}}\kappa \varphi =0
\label{Eq:H2}
\end{equation}%
\begin{equation}
-i\omega \mathbf{j}_{p,\parallel }+i\mathbf{q}s^{2}\delta n-i\mathbf{q}\frac{%
en_{0}}{m_{e}}\varphi =0\,,  \label{Eq:H3}
\end{equation}%
where $\mathbf{j}_{p,\parallel }=\left( j_{p,x},j_{p,y}\right) $ are the
in-plane components of the particle current density, and $%
\varphi $ and $\delta n$ are related by
\end{subequations}
\begin{equation}
\varphi =\frac{4\pi e\delta n}{\kappa ^{2}-q^{2}}  \label{Eq:H4}
\end{equation}%
through the Poisson equation. The set of equations has nontrivial solutions
only if
\begin{equation}
\kappa ^{2}=q^{2}+s^{-2}\left( \omega _{B}^{2}-\omega ^{2}\right) \,,
\label{InverseLocalizationLength}
\end{equation}%
where $\omega _{B}^{2}=\frac{4\pi e^{2}n_{0}}{m_e}$ is the bulk plasmon
frequency.

Now we can write down the general solution for the density oscillation as
\begin{equation}
\delta n=\delta n_{1}e^{\kappa z}e^{i\mathbf{q}\cdot \mathbf{r}}\,,~~~z<0
\label{Eq:dns-gen-sol}
\end{equation}%
and that for the electric potential as
\begin{eqnarray}
\varphi &=&\varphi _{1}e^{\kappa z}e^{i\mathbf{q}\cdot \mathbf{r}}+\varphi
_{2}e^{qz}e^{i\mathbf{q}\cdot \mathbf{r}}\text{, \ \ }z<0  \notag \\
\varphi &=&\varphi _{0}e^{-qz}e^{i\mathbf{q}\cdot \mathbf{r}}\text{, \ }z>0
\label{Eq:phi-gen-sol}
\end{eqnarray}%
with $\delta n_{1}$, $\varphi _{0}$, $\varphi _{1}$, and $\varphi _{2}$
being integration constants to be determined by the boundary conditions.

Putting the general solutions for the electron density fluctuation (\ref%
{Eq:dns-gen-sol}) and the electric potential~(\ref{Eq:phi-gen-sol}) into the
boundary conditions given by Eqs.~(\ref{Eq:bc-jz}) and (\ref{Eq:bc-phi}), we
obtain a set of three linear homogeneous equations, the solution of which
gives the dispersion relation%
\begin{equation}
\left( \kappa +q\right) \omega _{B}^{2}-2\kappa \omega ^{2}+\mathcal{P}%
\left( \kappa -q\right) \omega _{B}^{2}\omega \tau _{so}\sin \phi _{q}=0 \,,
\label{Eq: gen-3D-dispersion}
\end{equation}%
where $\phi _{q}$, with $-\pi <\phi _{q}\leq \pi $, is the angle between the
wave vector $\mathbf{q}$ and the in-plane magnetization, and the quantity $%
\tau _{so}\left( \equiv \frac{m_{e}\lambda ^{2}}{\hbar }\right) $ has the
dimension of time which characterizes the strength of the SO interaction.
Making use of Eq.~(\ref{InverseLocalizationLength}), one can rewrite Eq.~(%
\ref{Eq: gen-3D-dispersion}) as follows
\begin{equation}
\left( \kappa -q\right) \left[ \omega ^{2}-s^{2}\left( \kappa +q\right) ^{2}-%
\mathcal{P}\omega _{B}^{2}\omega \tau _{so}\sin \phi _{q}\right] =0 \,.
\label{Eq: gen-3D-dispersion1}
\end{equation}%
While the solution $\kappa =q$ gives the bulk frequency of $\omega =\omega
_{B}$ as can be easily seen from Eq.~(\ref{InverseLocalizationLength}), the
general surface plasmon frequency is given by the equation
\begin{equation}
\omega ^{2}-s^{2}\left( \kappa +q\right) ^{2}-\mathcal{P}\omega
_{B}^{2}\omega \tau _{so}\sin \phi _{q}=0 \,.  \label{Eq:gen-SP}
\end{equation}

It is instructive to first examine the solutions in the long wavelength
limit ($q\rightarrow 0$) for which Eq.~(\ref{Eq:gen-SP}) reduces to
\begin{equation}
2\omega ^{2}-\omega _{B}^{2}-\mathcal{P}\omega _{B}^{2}\omega \tau _{so}\sin
\phi _{q}=0 \,.  \label{Eq:SP-q=0}
\end{equation}%
Equation~(\ref{Eq:SP-q=0}) has two possible solutions
\begin{subequations}
\begin{equation}
\omega _{+}=\frac{\omega _{S}}{2}\left[ \mathcal{P}\omega _{S}\tau _{so}\sin
\phi _{q}+\sqrt{4+\left( \mathcal{P}\omega _{S}\tau _{so}\sin \phi
_{q}\right) ^{2}}\right]  \label{Eq:w+3D}
\end{equation}%
\begin{equation}
\omega _{-}=\frac{\omega _{S}}{2}\left[ \mathcal{P}\omega _{S}\tau _{so}\sin
\phi _{q}-\sqrt{4+\left( \mathcal{P}\omega _{S}\tau _{so}\sin \phi
_{q}\right) ^{2}}\right] \,,
\end{equation}%
where $\omega _{S}=\frac{\omega _{B}}{\sqrt{2}}$ is the 3D surface plasmon
frequency at $q=0$.
We observe that, for any angle $\phi _{q}$, the solution $\omega _{+}$ is
positive whereas the $\omega _{-}$ solution is negative. Moreover, the
solutions satisfy the relation $\omega _{+}\left( -\phi _{q}\right) =-\omega
_{-}\left( \phi _{q}\right) .$ The existence of two solutions connected in
this manner is a necessary condition for being able to construct real
solutions of the hydrodynamic equations. Physically, the two solutions
\textit{together} describe a single chiral wave whose frequency is
determined, for each wave vector $\mathbf{q}$ (specified by its magnitude $q$
and angle $\phi _{q}$) by the positive branch $\omega _{+}(\mathbf{q})$. A
real wave that propagates in the direction of $\mathbf{q}$ is described by
the superposition $e^{i\mathbf{q}\cdot \mathbf{r}}e^{-i\omega _{+}(\mathbf{q}%
)t}+e^{-i\mathbf{q}\cdot \mathbf{r}}e^{-i\omega _{-}(-\mathbf{q})t}$, where
we have used the fact that changing the sign of $\phi _{q}$ amounts to
reversing the direction of $\mathbf{q}$. Similarly, a real wave that
propagates in the direction of $-\mathbf{q}$ is described by the
superposition $e^{-i\mathbf{q}\cdot \mathbf{r}}e^{-i\omega _{+}(-\mathbf{q}%
)t}+e^{i\mathbf{q}\cdot \mathbf{r}}e^{-i\omega _{-}(\mathbf{q})t}$.
Crucially, the two waves, with wave vectors $\mathbf{q}$ and $-\mathbf{q}$
respectively, exhibit different phase velocities, and different dependences
on material parameters such as the strength of the SO interaction
characterized $\tau _{so}$, spin polarization $\mathcal{P}$, etc.

Another interesting feature of the 3D ferromagnetic surface plasmon is that
the decay length $\kappa ^{-1}$ behaves quite differently for waves that
propagate in opposite directions. More specifically, if the decay length $%
\kappa ^{-1}$ evaluated from Eq.~(\ref{InverseLocalizationLength}) with $%
\omega =\omega _{+}(\mathbf{q})$ \textit{increases }with increasing value of
$\left\vert \mathcal{P}\omega _{S}\tau _{so}\right\vert $ for surface
plasmons propagating along $\mathbf{q}$ direction, then it must \textit{%
decrease }with increasing value of $\left\vert \mathcal{P}\omega _{S}\tau
_{so}\right\vert $ for those propagating along $-\mathbf{q}$ direction. This
can be easily observed in the long wavelength limit for which $\omega_{+}$
is explicitly given by Eq.~(\ref{Eq:w+3D}). Furthermore, we note that when
the product $\left\vert \mathcal{P}\omega _{S}\tau _{so}\right\vert $
becomes sufficiently large, the surface plasmon mode is forbidden in a range
of directions where $\kappa $ remains imaginary (physically this means that
the surface mode merges with the bulk mode in these directions). For the
long wavelength limit, one can show that when $\left\vert \mathcal{P}\omega
_{S}\tau _{so}\right\vert >\frac{1}{\sqrt{2}}$, there exist two intervals
for the plasmon propagation angle $\phi _{q}\in \left[ \phi _{q}^{crit}-\pi
,-\phi _{q}^{crit}\right] \cup \left[ \phi _{q}^{crit},\pi -\phi _{q}^{crit}%
\right] $ with $\phi _{q}^{crit}=\arcsin \left( \frac{1}{\sqrt{2}\left\vert
\mathcal{P}\omega _{S}\tau _{so}\right\vert }\right) $, in which surface
mode is absent, as shown schematically in Fig.~\ref{Fig:w3D-v-q}(d).


\begin{figure*}[ht]
\centering
\includegraphics[trim={1cm 6cm 1cm
6cm},clip=true,width=\textwidth]{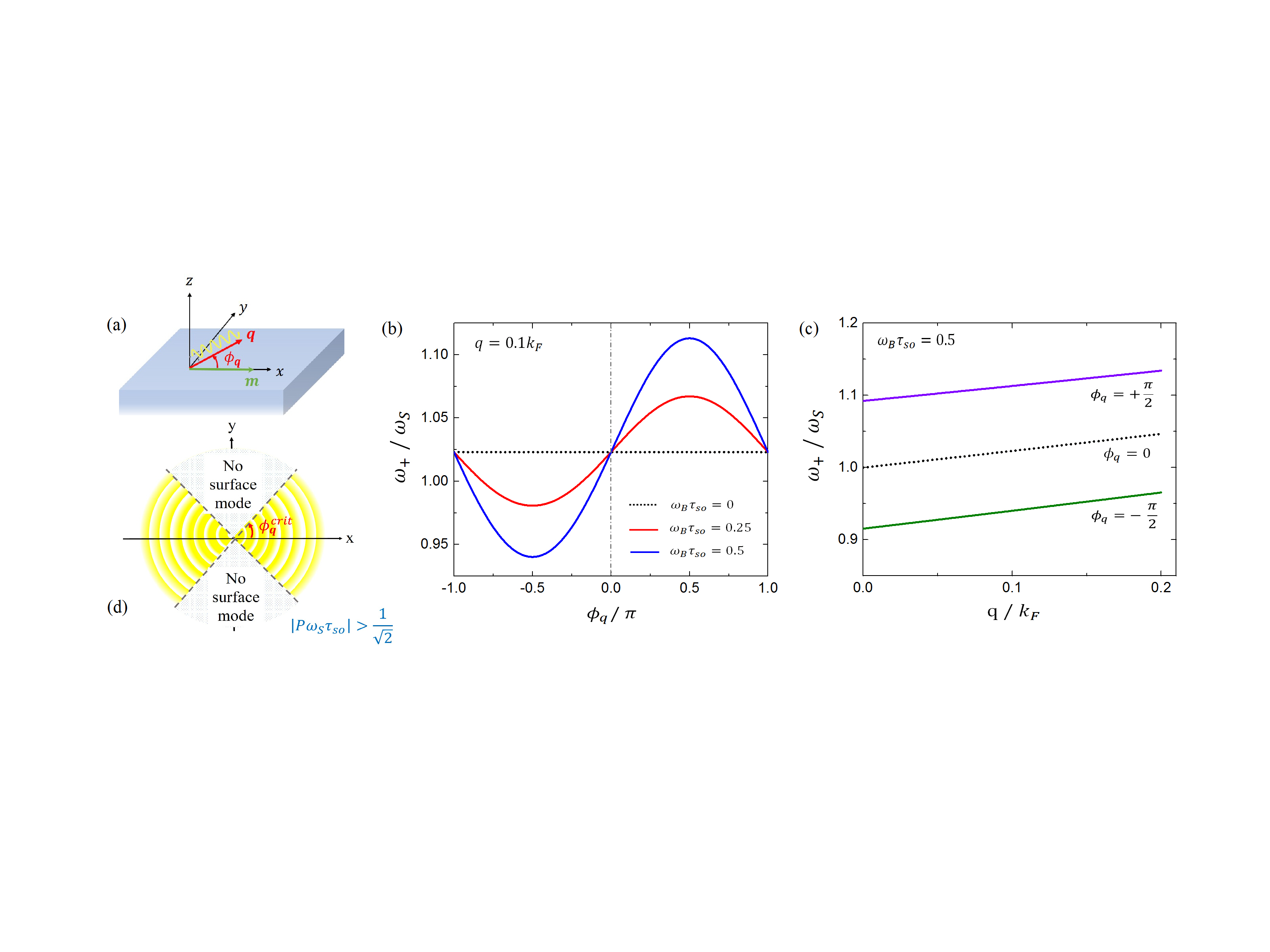}
\caption{Variation of the 3D ferromagnetic surface plasmon frequency $%
\protect\omega_{+}$ (scaled by $\protect\omega_S$) with the in-plane wave
vector $\mathbf{q}$. The set-up of the system is shown schematically in
panel (a) with $\protect\phi_q$ defined as the angle between the direction
of propagation of the surface plasmon and the direction of magnetization
fixed on the $x$-axis. Panel (b) shows $\protect\omega_{+}$ as a function of
$\protect\phi_q$ at a fixed magnitude of the wave vector of $q=0.1k_F$ for
several different values of $\protect\tau_{so}$ (scaled by $\protect\omega%
_B^{-1}$), panel (c) shows $\protect\omega_{+}$ as a function of the
magnitude of the wave vector $q$ for several different angles $\protect\phi%
_q $ with a given SO interaction strength $\protect\omega_B\protect\tau%
_{so}=0.5$, and panel (d) is a schematic picture showing the range of $%
\protect\phi_q$ (plasmon propagation directions) for which the surface mode
is forbidden when the quantity $\left\vert \mathcal{P}\protect\omega _{S}%
\protect\tau _{so}\right\vert$ is greater than $\frac{1}{\protect\sqrt{2}}$.}
\label{Fig:w3D-v-q}
\end{figure*}

In Fig.~\ref{Fig:w3D-v-q}, we show the ferromagnetic surface plasmon
frequency as a function of the direction and the magnitude of the wave
vector $\mathbf{q}$. For a given $q\left(=0.1k_{F}\right) $, the surface
plasmon frequency exhibits a sinusoidal-like dependence on $\phi _{q}$ as
shown in Fig.~\ref{Fig:w3D-v-q}(b): It reaches a maximum at $\phi _{q}=+%
\frac{\pi }{2} $ and a minimum at $\phi _{q}=-\frac{\pi }{2}$ (i.e., when
the surface plasmon propagates in directions perpendicular to the
magnetization), and coincides with the normal surface plasmon frequency in
the absence of SO interaction (indicated by the black dotted line) when the
surface plasmon propagates parallel or antiparallel to the magnetization
(i.e., $\phi _{q}=0$ or $\pi $).

Fig.~\ref{Fig:w3D-v-q}(c) shows the surface plasmon frequency as a function
of the magnitude of the plasmon wave vector for three different directions
of propagation, $\phi_q=-\frac{\pi}{2},0,\frac{\pi}{2}$. We note that while $%
\omega_{+}$ grows monotonically with $q$ for all three directions, their
frequencies remain non-degenerate for any $q$ due to the presence of the SO
interaction. One interesting consequence of the anisotropic dispersion is
that the phase velocities, given by $\frac{\omega_{+}(\mathbf{q})}{q}$, for
surface plasmons propagating in opposite directions ($\phi_q \neq 0,\pi $)
are always different: this implies that as long as the direction of
propagation deviates from the direction of the magnetization, no standing
wave can be formed for surface plasmons. 

\begin{figure}[ht]
\centering
\includegraphics[trim={6cm 3.5cm 4cm
3cm},clip=true,width=0.9\columnwidth]{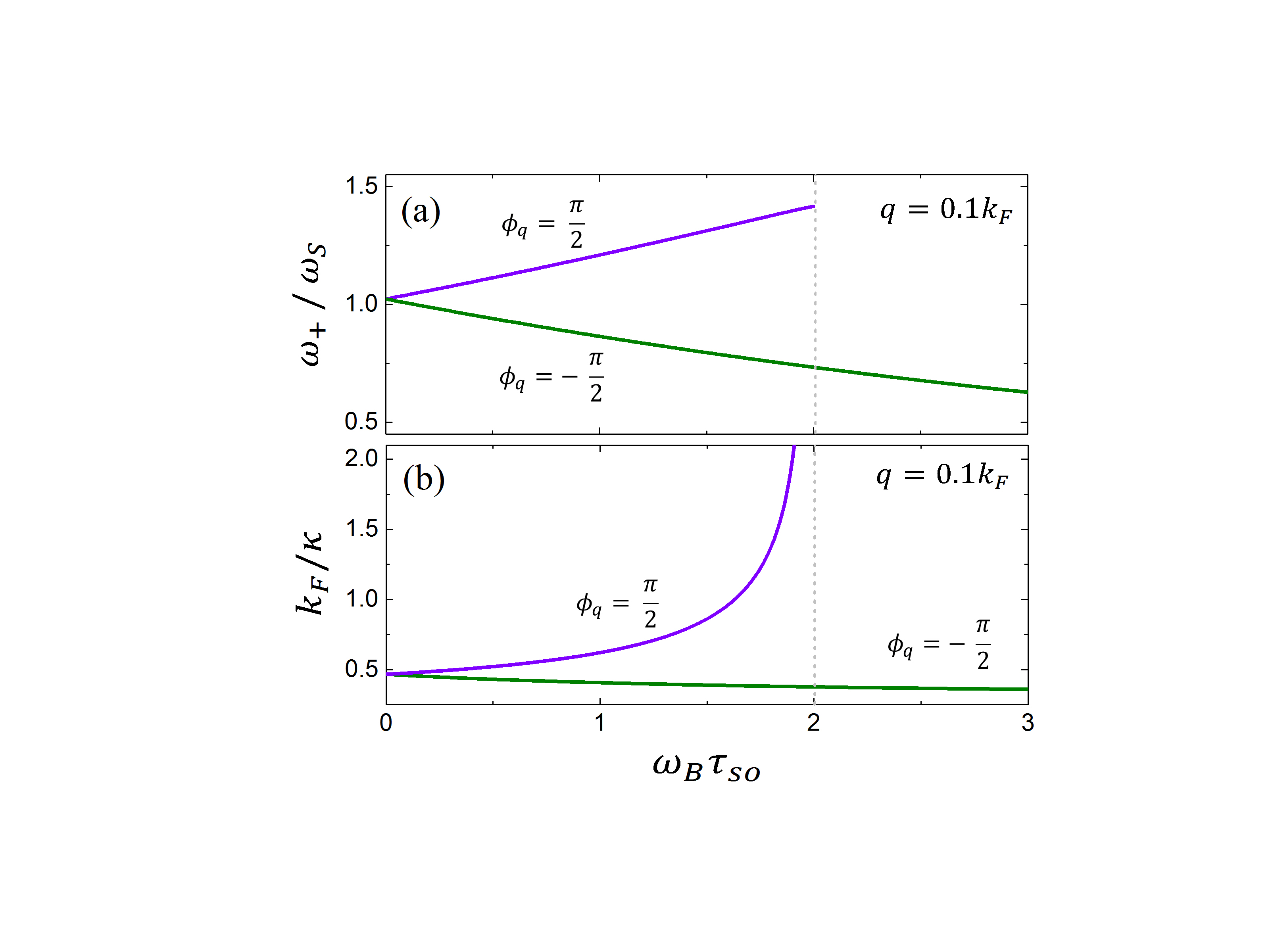}
\caption{Dependence of 3D ferromagnetic surface plasmon on the strength of
the SOC (characterized by $\protect\omega_B \protect\tau_{so}$): (a) The
ferromagnetic surface plasmon frequency $\protect\omega$ (scaled by $\protect%
\omega_S$) as a function of $\protect\omega_B\protect\tau_{so}$ and (b) the
decay length of the surface mode given by $\protect\kappa^{-1}$ as a
function of $\protect\omega_B\protect\tau_{so}$ at a given magnitude of wave
vector $q=0.1k_F$, including two opposite propagation directions: $%
\protect\phi_q=\frac{\protect\pi}{2}$ and $-\frac{\protect\pi}{2}$
respectively. }
\label{Fig:w3D-v-tauSO}
\end{figure}

In Fig.~\ref{Fig:w3D-v-tauSO}, we show the dependences of the frequency $\omega_{+}$ and
decaying length $\kappa^{-1}$ of the surface plasmons on the strength of the
SO interaction characterized by $\omega_B\tau_{so}$. As the effect of the SO
interaction is most prominent for surface plasmons propagating in the
directions perpendicular to the magnetization, we shall focus on the cases
of $\phi_q=-\frac{\pi}{2}$ and $\frac{\pi}{2}$ for a finite magnitude of the
wave vector $q$. Consistent with the qualitative analysis we performed for
the long wave length limit, we find that for surface plasmons propagating at
an angle $\phi_q=-\frac{\pi}{2}$ with respect to the magnetization, both $%
\omega_{+}$ and $\kappa^{-1}$ decrease monotonically with increasing $%
\omega_B\tau_{so}$, whereas for those propagating in the opposite direction
(i.e., $\phi_q=\frac{\pi}{2}$) both $\omega_{+}$ and $\kappa^{-1}$ increase
monotonically with increasing $\omega_B\tau_{so}$ and are terminated when $%
\tau_{so}$ reaches a certain threshold (indicated by the vertical dashed
line in the Fig.~\ref{Fig:w3D-v-tauSO}) where the decay length $\kappa^{-1}$ diverges as
$\omega^2$ approaches $\omega_B^2+s^2q^2$ (cf. Eq.~(\ref%
{InverseLocalizationLength})) and the surface mode merges into the bulk
mode~\footnote{%
A positive solution to the surface plasmon dispersion equation~(\ref%
{Eq:gen-SP}) does not exist when the magnitude of $\omega_B\tau_{so}$ is
further increased beyond the threshold, leaving the bulk plasmon frequency
as the only solution to the general plasmon dispersion equation~(\ref{Eq:
gen-3D-dispersion}) or (\ref{Eq: gen-3D-dispersion1}))}.

\section{Solution in two dimensions}
\label{sec:2D-sol}

An exact treatment of the electrostatics in a two-dimensional plane is quite
more complicated than in 3D, due to the fact that the electric field exists
in the whole three-dimensional space, while the electron density is confined
to a plane. Fortunately, the treatment can be greatly simplified by making
the approximation adopted by Fetter~\cite{Fetter85} in his treatment of the
two-dimensional edge magnetoplasmon, namely replacing the exact
electrostatic Green's function (the nonlocal kernel that connects the
density to the potential in the plane) by an approximate Green's function
that has the same integrated area and second moment. What is lost in the
approximation is a weak logarithmic dependence of the edge magnetoplasmon
frequency on the magnitude of the wave vector, $\omega \sim q|\ln q|$, which
is confirmed by a more accurate treatment making use of the Wiener-Hopf
technique.~\cite{Volkov85,Noble58} This is not a very serious drawback in
our case, since the long wavelength dispersion continues to be largely
controlled by classical electrostatics, which mandates a $\sqrt{q}$
dependence. With this approximation, the equations remain essentially the
same as in 3D, except that Eq.~(\ref{Eq:w2D-gen1}), connecting the potential
to the density, takes the slightly different form
\end{subequations}
\begin{equation}
\varphi =\frac{4\pi e|q|}{\kappa ^{2}-2q^{2}}\delta n\,.  \label{Eq:phi_2d}
\end{equation}%
\thinspace We let the electron liquid be confined to the $y$-$z$ plane, and
consider the edge plasmon localized in the $z$-direction and propagating in
the $y$-direction, while the magnetization is along the $x$ axis,
perpendicular to the electron liquid (since the plasmon only propagates in
the $y$-direction, we have suppressed the subscript for the wave vector $q.$%
). Combining Eq.~(\ref{Eq:phi_2d}) with the set of equations (\ref{Eq:H1}-%
\ref{Eq:H3}) (with $\mathbf{q}$ replaced by $q\mathbf{\hat{y}}$), we arrive
at the equation relating the edge plasmon frequency $\omega $ and the
decaying constant $\kappa $, i.e.,
\begin{equation}
\kappa ^{4}-\kappa ^{2}\left( k_{0}^{2}+k_{\omega }^{2}+3q^{2}\right)
+2q^{2}\left( k_{\omega }^{2}+q^{2}\right) =0\,,
\end{equation}%
where we have defined
\begin{equation}
k_{\omega }^{2}\equiv \frac{\omega _{q}^{2}-\omega ^{2}}{s^{2}}\text{ and }%
k_{0}^{2}\equiv \frac{\omega _{q}^{2}}{s^{2}}\,
\end{equation}%
with $\omega _{q}=\sqrt{\frac{2\pi n_{0}e^{2}\left\vert q\right\vert }{m_{e}}%
}$ the bulk 2D plasmon frequency. The equation has two solutions:%
\begin{eqnarray}
\kappa _{1,2}^{2} &=&\frac{1}{2}\left[ k_{0}^{2}+k_{\omega
}^{2}+3q^{2}\right.  \notag \\
&&\left. \pm \sqrt{\left( k_{0}^{2}+k_{\omega }^{2}\right) ^{2}+q^{2}\left(
6k_{0}^{2}-2k_{\omega }^{2}+q^{2}\right) }\right] \,.  \label{Eq:kappa12}
\end{eqnarray}%
Similar to the 3D case, we write the general solutions for the electrostatic
potential and the electron density fluctuation as follows

\begin{eqnarray}
\varphi &=&\varphi _{1}e^{\kappa _{1}z}e^{iqy}+\varphi _{2}e^{\kappa
_{2}z}e^{iqy}\text{, \ \ }z<0  \notag \\
\varphi &=&\varphi _{0}e^{-\sqrt{2}qz}e^{iqy}\text{, \ }z>0
\end{eqnarray}%
\begin{equation}
\delta n=\frac{1}{4\pi e\left\vert q \right\vert}\left[ \left( \kappa _{1}^{2}-2q^{2}\right)
\varphi _{1}e^{\kappa _{1}z}+\left( \kappa _{2}^{2}-2q^{2}\right) \varphi
_{2}e^{\kappa _{2}z}\right] e^{iqy}\,,
\end{equation}%
where we have suppressed the common time-dependent components. The general
solution for the density fluctuation $\delta n$ was derived by invoking the
approximate Poisson relation~(\ref{Eq:phi_2d}). Putting these equations in
the boundary conditions, we derive the following equation for $\omega $, the
solution of which gives the edge plasmon frequency, i.e.,%
\begin{eqnarray}
\frac{\kappa _{1}\left( \kappa _{1}^{2}-2q^{2}-2k_{0}^{2}\right) +2\mathcal{P%
}k_{0}^{2}\omega \tau _{so}qm_{x}}{\kappa _{2}\left( \kappa
_{2}^{2}-2q^{2}-2k_{0}^{2}\right) +2\mathcal{P}k_{0}^{2}\omega \tau
_{so}qm_{x}} &=&\frac{\kappa _{1}+\sqrt{2}\left\vert q\right\vert }{\kappa
_{2}+\sqrt{2}\left\vert q\right\vert }\,.  \notag \\
&&  \label{Eq:w2D-gen1}
\end{eqnarray}%
Note that the magnetization enters the dispersion relation only through the $%
x$-component of the magnetization (i.e., the one perpendicular to the plane
of the 2D electron liquid). This can be understood as follows: The
magnetization enters the formula for the anomalous velocity as $\mathbf{v}%
_{a}\sim \mathbf{\hat{m}\times }\nabla \varphi $; as the surface plasmon
propagates along the edge in the $y$-direction, only the $x$-component of
the magnetization generates an anomalous velocity in the $z$-direction
(perpendicular to the edge), which affects the boundary condition and hence
the edge plasmon frequency.

In the long wavelength limit, we obtain simpler forms for $\kappa _{1}$ and $%
\kappa _{2}$ by keeping only leading order terms in $q$, i.e.,
\begin{equation}
\kappa _{1}\simeq C_{\omega }\left\vert q\right\vert +O\left( q^{3}\right)
\end{equation}%
with $C_{\omega }^{2}=\frac{\omega _{q}^{2}-\omega ^{2}}{\omega _{q}^{2}-%
\frac{1}{2}\omega ^{2}}$ and%
\begin{equation}
\kappa _{2}\simeq \frac{1}{s}\sqrt{2\omega _{q}^{2}-\omega ^{2}}+O\left(
q^{2}\right) \,.
\end{equation}%
Making use of these expressions, the general equation~(\ref{Eq:w2D-gen1})
can be reduced to
\begin{equation}
\omega ^{2}-\frac{2}{3}\omega _{q}^{2}+\frac{2\sqrt{2}}{3}\mathcal{P}\omega
_{q}^{2}\omega \tau _{so}m_{x}=0 \,,
\end{equation}%
where we have made the approximation of $C_{\omega }\simeq \frac{\sqrt{2}}{2}
$ by taking $\omega $ to be the unperturbed edge plasmon frequency $\omega
^{\left( 0\right) }=\sqrt{\frac{2}{3}}\omega _{q}$.~\cite{Fetter85} This
equation has two solutions,
\begin{equation}
\omega _{+}=\frac{\sqrt{2}}{3}\omega _{q}\left[ \sqrt{3+\left( \mathcal{P}%
\omega _{q}\tau _{so}m_{x}\right) ^{2}}-\mathcal{P}\omega _{q}\tau _{so}m_{x}%
\mathrm{sgn}\left( q\right) \right]
\end{equation}%
and
\begin{equation}
\omega _{-}=-\frac{\sqrt{2}}{3}\omega _{q}\left[ \sqrt{3+\left( \mathcal{P}%
\omega _{q}\tau _{so}m_{x}\right) ^{2}}+\mathcal{P}\omega _{q}\tau _{so}m_{x}%
\mathrm{sgn}\left( q\right) \right] \,.
\end{equation}%
Notice that in the absence of the SO interaction, we recover the unperturbed
edge plasmon frequency. As in the 3D case, the reality of the classical wave
fields requires $\omega _{+}\left( -q\right) =-\omega _{-}\left( q\right) $,
\textit{which shows the solutions of opposite frequencies and momenta to be
parts of the same wave}. Also, reversing the direction of propagation of the
edge plasmon is equivalent to reversing the direction of the magnetization
direction: therefore we find $\omega _{+}\left( -\mathbf{m}\right) =-\omega
_{-}\left( \mathbf{m}\right) $ as expected. For this reason, we shall
concentrate on the positive solution $\omega_{+}$ of the 2D ferromagnetic
edge plasmons in the following discussions.

Properties of the 2D ferromagnetic edge plasmon at finite wavelength are
readily obtained by numerically solving Eq.~(\ref{Eq:w2D-gen1}). In Fig.~\ref%
{Fig:w2D-v-q_thetaM}(a), we plot $\omega _{+}$ as a function of wave vector $%
q$ with the magnetization direction fixed along the $x$-axis. In the absence
of the SO interaction (i.e., $\omega_{k_F}\tau _{so}=0$), $\omega _{+}$ is
symmetric in $q$ as indicated by the black dotted lines. The frequency of
the left-propagating mode increases with increasing strength of the SO
interaction, whereas that of the right-propagating mode decreases. In
addition, we observe that the left- and right-propagating modes remain
gapless at $q=0$, in contrast to the 2D edge magnetoplasmon~\cite{Fetter85},
which develops a gap in one direction. We will come back to this point in
the next section when we compare the ferromagnetic surface plasmon with the
more familiar surface magnetoplasmon. A similar chiral plasmon dispersion
was also found in massive Dirac systems~\cite{Low2015}.

\begin{figure}[h]
\includegraphics[trim={7cm 0cm 8cm 2.0cm},clip=true,
width=0.9\columnwidth]{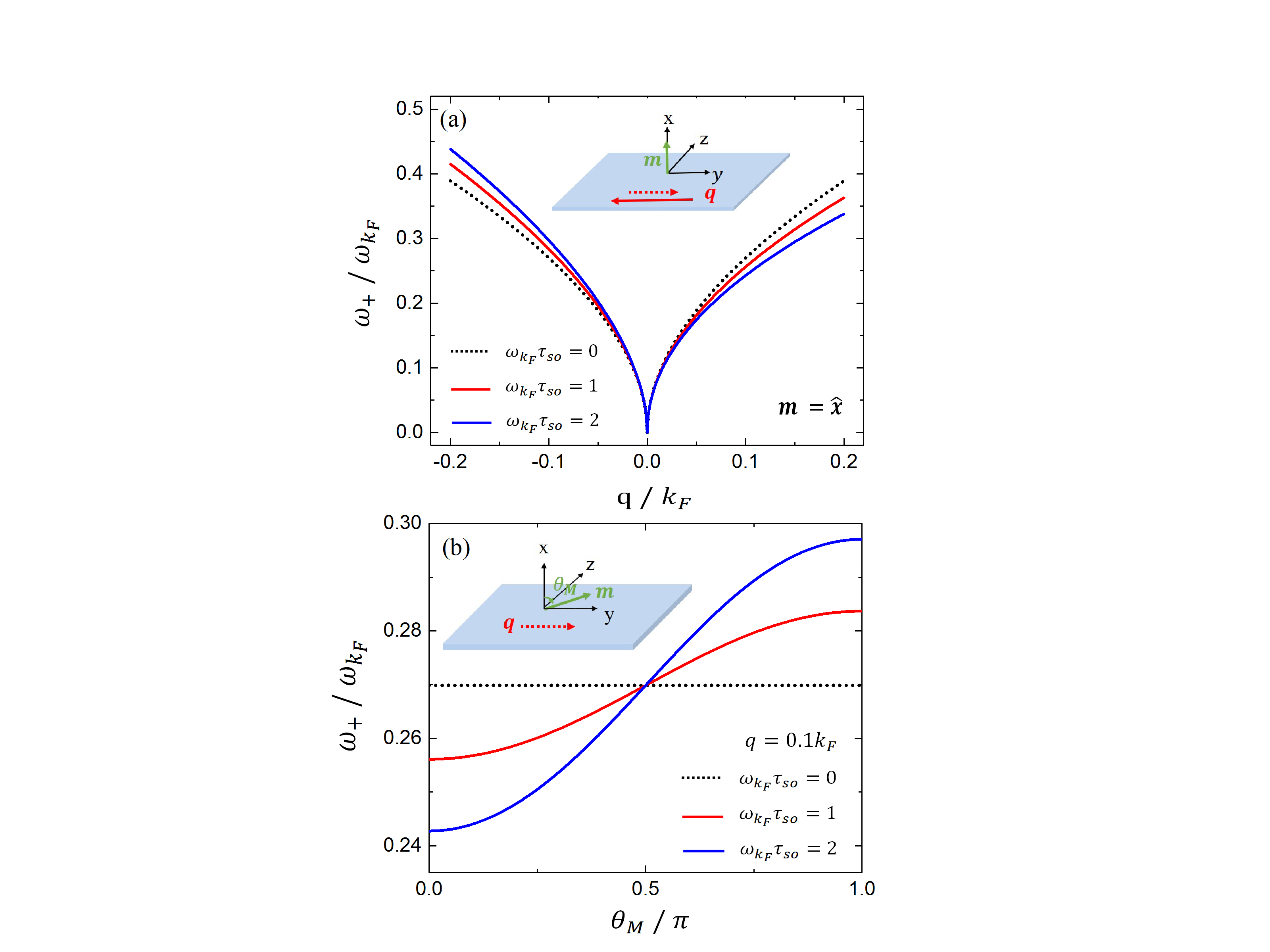}
\caption{Dependence of the 2D ferromagnetic edge plasmon frequency (scaled
with bulk plasmon frequency with $q=k_F$) on (a) the wave vector along the $%
y $-direction with magnetization direction (denoted by $\mathbf{m}$) fixed
in the $x$-direction and (b) angle $\protect\theta_M$ between the
magnetization $\mathbf{m}$ and the $x$-direction (i.e., the normal direction
of 2D electron liquid plane), including three different strengths of the SO
interaction. }
\label{Fig:w2D-v-q_thetaM}
\end{figure}

In Fig.~\ref{Fig:w2D-v-q_thetaM}(b), we plot $\omega _{+}$ as a function of
the polar angle $\theta _{M}$ between the magnetization and the $x$-axis for
a right-propagating wave with a given wave vector of $q=0.1 k_{F}$ where $%
k_{F}$ is the Fermi wavelength. The extremes in $\omega _{+}$ occur when the
magnetization is perpendicular to the plane of the 2D electron fluid with a
maximum at $\theta _{M}=\pi $ and a minimum at $\theta _{M}=0$. The SO
interaction has no effect on the edge ferromagnetic plasmon when the
magnetization lies in the plane of the 2D electron fluid as shown by the
crossing point at $\theta _{M}=\frac{\pi }{2}$.

\begin{figure}[h]
\includegraphics[trim={5.2cm 4.5cm 2.8cm 1.8cm},clip=true,
width=0.9\columnwidth]{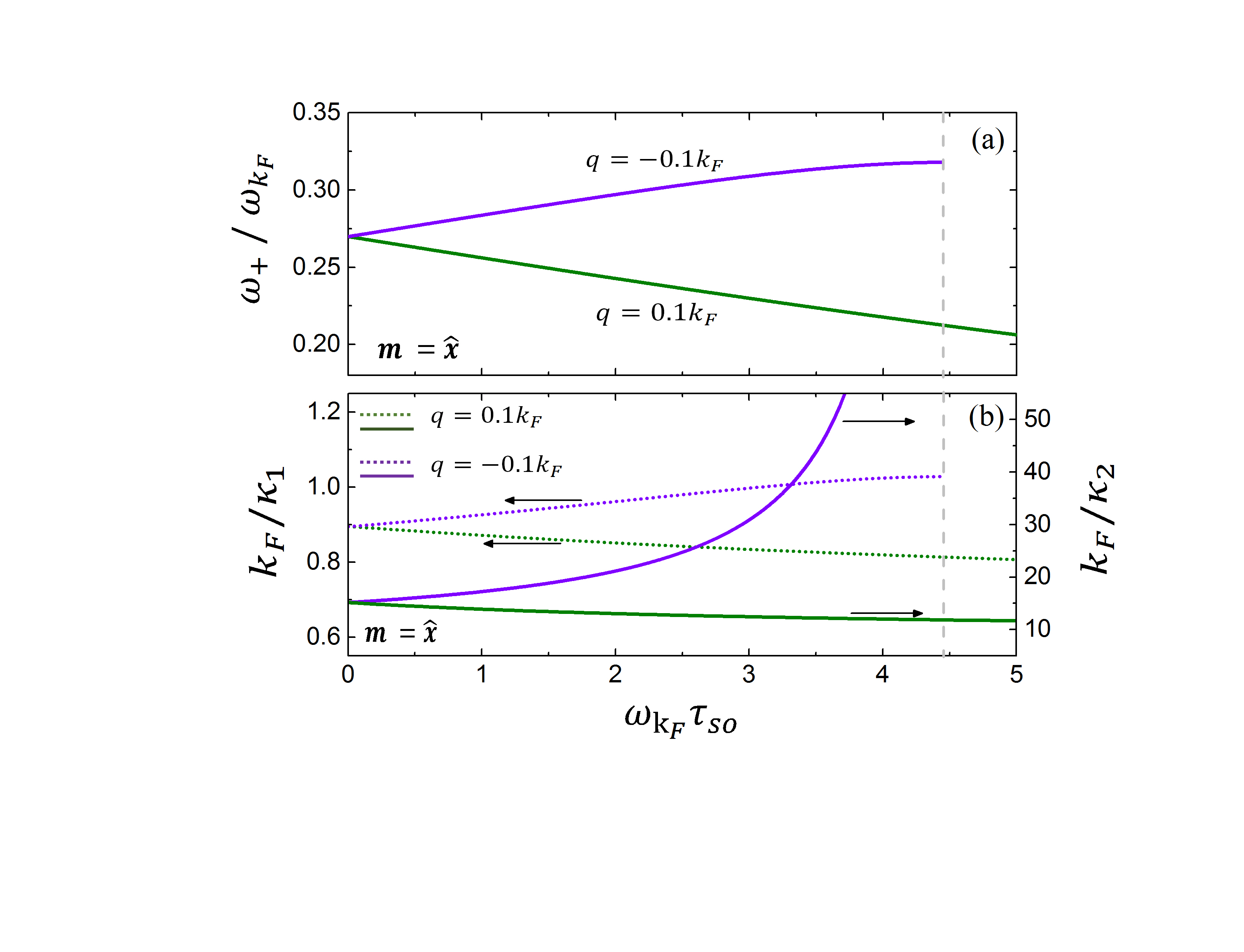}
\caption{(a) Variation of the edge ferromagnetic surface plasmon frequency $%
\protect\omega$ vs SO interaction strength parameter $\protect\omega_{k_F}%
\protect\tau_{so}$ and (b) variation of the decay lengths of the edge modes $%
\protect\kappa _{1,2}^{-1}$ vs $\protect\omega_{k_F}\protect\tau_{so}$. Two
opposite wave vectors $q=0.1k_F$ and $q=-0.1k_F$ are considered. The
vertical grey dashed lines indicate the critical magnitude of the SO
interaction beyond which the edge mode with $q=-0.1k_F$ no longer exists.}
\label{Fig:w2D-v-tauSO}
\end{figure}

Lastly in Fig.~\ref{Fig:w2D-v-tauSO}, we show the variations of the
frequency $\omega _{+}$ and decaying lengths $\kappa _{1}^{-1}$ and $\kappa
_{2}^{-1}$, given by Eq.~(\ref{Eq:kappa12}) ), as functions of the strength
of the SO interaction. For the right-propagating mode (i.e., $q=0.1k_{F}$),
both $\omega _{+}$ and $\kappa _{i}^{-1}$ ($i=1,2$) decrease with increasing
$\omega_{k_F}\tau _{so}$, whereas for left-propagating mode (i.e., $%
q=-0.1k_{F}$) both $\omega _{+}$ and $\kappa _{i}^{-1}$ ($i=1,2$) increase
monotonically with increasing $\omega_{k_F}\tau _{so}$ and terminate,
similar to the 3D case, when $\omega_{k_F}\tau _{so}$ reaches a threshold
(as indicated by the vertical dashed line) beyond which the edge mode merges
with the bulk mode.

\section{Discussion}
\label{Sec:discussion}

\subsection{Comparison with surface and edge magnetoplasmon}

Now that we have thoroughly investigated both the surface and edge plasmons
in 3D and 2D ferromagnetic conductors, it is worthwhile discussing the
features that distinguish them from the classical surface or edge
magnetoplasmons. In the classic magnetoplasmon, the direction dependence of
the plasmon dispersion arises from the Lorentz force exerted by the applied
magnetic field, while in the present case there is no magnetic field, but an
anomalous velocity connecting the collective charge oscillation with the
bulk magnetization. In addition, we note that the anomalous velocity term
plays a role in altering surface ferromagnetic plasmon frequency only
through the boundary conditions, whereas the Lorentz force contributes to
the time rate of change of the canonical current (or momentum) density and
hence enters the bulk Euler equation~\cite{Fetter85}. These essential
differences are reflected in the dispersion relations. Qualitative
differences emerge already in the long wavelength limit, as we show below.

For a surface magnetoplasmon that propagates in the $(x,y)$ plane, with
magnetic field lying in the same plane, the $q\rightarrow 0$ limit of the
dispersion is given by
\begin{equation}
\omega _{mp}^{3D}(\phi _{q})=\frac{\omega _{c}\sin \phi _{q}}{2}+\sqrt{\frac{%
\omega _{c}^{2}\sin ^{2}\phi _{q}}{4}+\frac{\omega _{B}^{2}+\omega
_{c}^{2}\cos ^{2}\phi _{q}}{2}}\,,  \label{Eq:gen-wMP3D}
\end{equation}%
where $\omega _{c}\left( >0\right) $ is the cyclotron frequency and $\phi
_{q}$ is the angle between $\mathbf{q}$ and the applied magnetic field. A
detailed derivation of surface magnetoplasmon dispersion with arbitrary
propagation direction is presented in Appendix A. For plasmons propagating
perpendicular to the in-plane magnetic field, i.e., $\phi _{q}=\pm \frac{\pi
}{2}$, Eq.~(\ref{Eq:gen-wMP3D}) reduces to

\begin{equation}
\omega _{mp,\perp }^{3D}=\frac{1}{2}\left( \sqrt{\omega _{c}^{2}+2\omega
_{B}^{2}}\pm \omega _{c}\right) \,.
\end{equation}%
which is exactly the result for the special case discussed by Fetter~\cite%
{Fetter85}. The general dispersion (\ref{Eq:gen-wMP3D}) also shows that even
when the plasmon wave vector is collinear with the magnetic field (i.e.,
when $\phi_q=0$), the magnetic field still gives rise to a correction to the
surface magnetoplasmon frequency of second order in $\omega _{c}$, i.e.,
\begin{equation}
\omega _{mp,\parallel }^{3D}=\sqrt{\frac{\omega _{B}^{2}+\omega _{c}^{2}}{2}}%
\,,
\end{equation}%
This may seem a little counterintuitive at first glance as one may think the
Lorentz force, given by $\frac{e}{c}\mathbf{j\times H}\ $with $\mathbf{H}$
and $\mathbf{j}$ the magnetic field and the current density respectively,
would vanish in this geometry; however, this is in fact not the case since
the in-plane current density is in general \textit{not }parallel to the wave
vector in the presence of the magnetic field (see Appendix A for the general
relation between $\mathbf{j}$ and $\mathbf{q}$: the component of the
electric field perpendicular to the surface combines with the magnetic field
to produce a drift velocity perpendicular to the magnetic field). In
addition, due to mirror symmetry about the plane perpendicular to the
magnetic field, the frequencies of surface plasmons propagating parallel or
antiparallel to the applied magnetic field must be identical, giving rise to
an effect of the order of $O\left( \omega _{c}^{2}\right) $. Although the
frequency of the ferromagnetic surface plasmons also remains finite at $q=0$%
, it reduces to the normal surface plasmon frequency of $\omega _{S}=\frac{%
\omega _{B}}{\sqrt{2}}$ when the propagation direction of the plasmon waves
become collinear with the in-plane magnetization (i.e., $\phi _{q}=0$ or $%
\pi $). The reason for this different behavior is that the anomalous
velocity ceases to be operative in the boundary condition for the current
density when $\mathbf{q}$ is parallel to $\hat{\mathbf{m}}$ (see Eq.~(\ref%
{Eq:bc-jz}). Another difference is that surface magnetoplasmons in the long
wavelength limit remain well defined for all directions of $\mathbf{q}$ and
all values of the magnetic field, at variance with ferromagnetic plasmons
which in certain directions may merge with the bulk plasmons when the SO
interaction is strong enough.

\begin{figure}[h]
\includegraphics[trim={1cm 0cm 2cm 2.0cm},clip=true, width=0.9
\columnwidth]{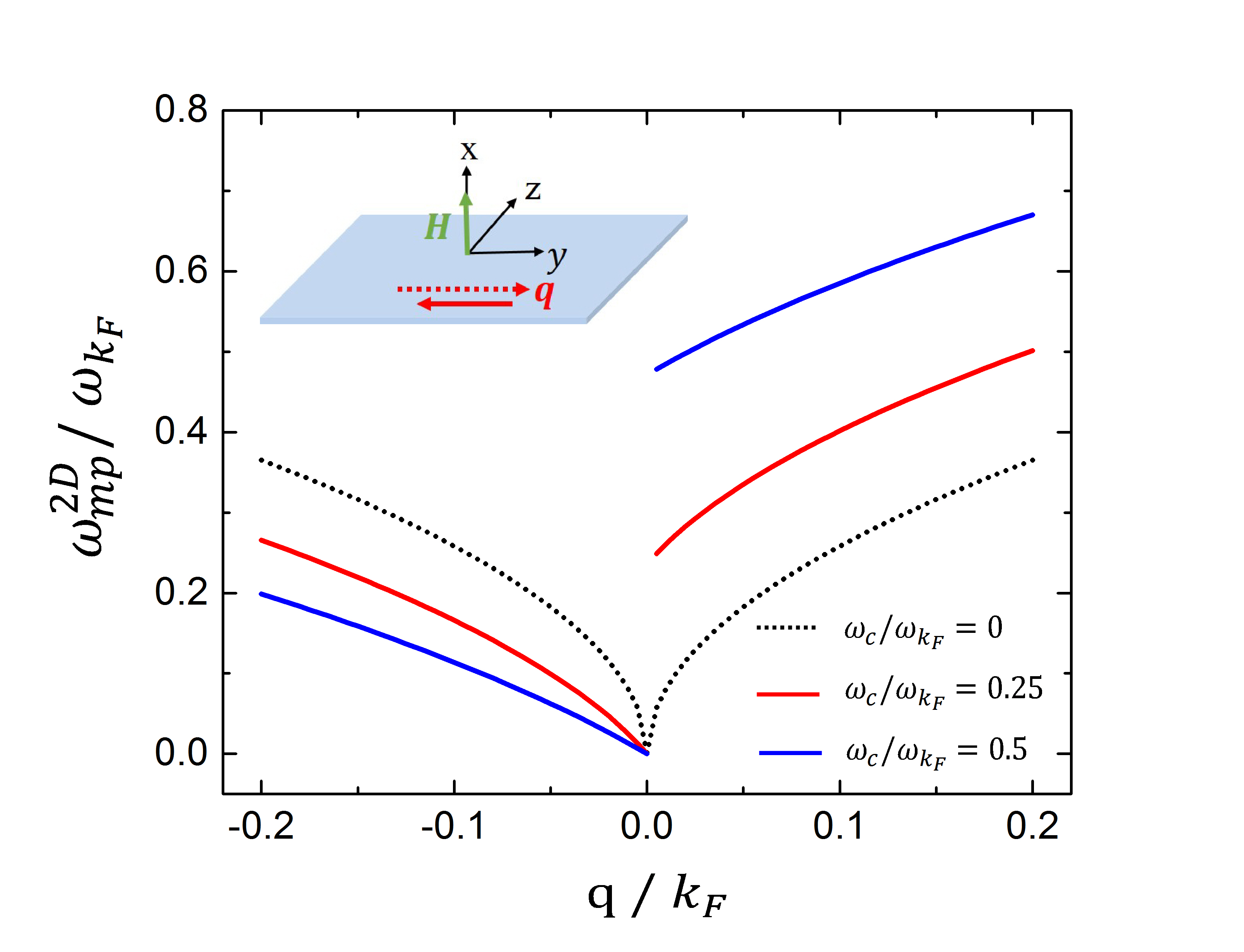}
\caption{2D edge magnetoplasmon dispersion relation with the external
magnetic field (denoted by $\mathbf{H}$) fixed in the $x$-direction, for
three different magnitudes of $\mathbf{H}$. }
\label{Fig:w2D-v-q_Fetter}
\end{figure}

More significant differences arise in two dimensions. Quoting from Ref.~%
\onlinecite{Fetter85} the dispersion of the low-frequency edge
magnetoplasmon for $q\rightarrow 0$ is
\begin{equation}
\omega _{mp}^{2D}=\frac{\sqrt{2}}{3}\left[ \sqrt{3\omega _{q}^{2}+\omega
_{c}^{2}}+\omega _{c}sgn\left( q\right) \right] ,
\end{equation}%
where $\omega _{c}>0$ is the cyclotron frequency, and $\omega _{q}\left( =%
\sqrt{\frac{2\pi n_{0}e^{2}\left\vert q\right\vert }{m_{e}}}\right) $ is the
bulk 2D plasmon frequency. We see that in this case the right-propagating
mode with $q>0$ is gapped and approaches a frequency $\frac{2\sqrt{2}}{3}%
\omega _{c}$ in the long wavelength limit, whereas the left-propagating mode
with $q<0$ has a frequency that goes to zero linearly for $q\rightarrow 0$,
as shown in Fig.~\ref{Fig:w2D-v-q_Fetter}. This is quite different from the
ferromagnetic edge plasmons, where both right- and left propagating waves
have a dispersion $\omega \propto \omega _{q}\propto \sqrt{q}$ for $%
q\rightarrow 0$, only with different proportionality constants in the two
directions, as shown by Fig.~\ref{Fig:w2D-v-q_thetaM}(a).

\subsection{Material considerations}

In transition metal ferromagnets, contributions to the anomalous Hall effect
due to the intrinsic or side jump mechanism can be attributed to an
anomalous velocity~\cite{Nagaosa10RMP} which, in the present paper, has been
shown to play an essential role in imparting chiral properties to the
ferromagnetic surface plasmons. Consequently, one would expect the
ferromagnetic surface plasmons to be observable in transition metal
ferromagnets with large anomalous Hall effect. To obtain some order of
magnitude estimate for the anisotropy of the ferromagnetic surface plasmon
frequency in the long wavelength limit, as parameterized by $\eta _{S}\equiv
\frac{\omega \left(\phi_q=\frac{\pi }{2}\right) -\omega
\left(\phi_q=0\right) }{\omega \left(\phi_q=0\right) }$, let us consider the
anomalous Hall conductivity due to side jump, which is given by $\sigma
_{yx}^{sj}\sim \frac{n_{0}e^{2}}{\hbar }\lambda ^{2}$.~\cite%
{Holstein72PRL,Berger70PRB_sidejump,Vignale10} The effective Compton
wavelength $\lambda $ can thus be estimated from the experimentally
accessible anomalous Hall angle $\theta _{ah}$ ($\equiv \frac{ \sigma _{yx}}{%
\sigma _{xx}},\text{with }\sigma _{xx}=\frac{e^{2}n_{0}\tau }{m_{e}}$ the
longitudinal conductivity) by $\tau _{so}=\frac{m_{e}\lambda ^{2}}{\hbar }%
=\theta _{ah}\tau $ where $\tau $ is the momentum relaxation time. Taking
the parameters $\tau \sim 10^{-14}\,$s, $\theta _{ah}=0.01$, $\mathcal{P}%
=0.5 $, $\omega _{B}\sim 50$ THz~\cite{Ordal83ApplOpt_plasm}, we find $%
\tau_{so}=0.1$~fs and $\omega_B\tau_{so}=0.005$ which lead to $\eta _{S}\sim
0.1\%$.

Another promising class of materials to observe the 3D ferromagnetic surface
plasmons are the diluted magnetic semiconductors~\cite{Dietl03NatMater_DMS}.
For example, using the following material parameters for GaMnAs: $\omega
_{B}\sim 100$ THz, $\lambda ^{2}\simeq 4.4$ \AA $^{2}$, $m_{e}\sim
10^{-31}\, $kg,~\cite{Matz81PRL_plasmon-GaAs,Winkler03} we find $%
\tau_{so}=0.01$~fs and $\omega_B\tau_{so}=0.001$ for which the parameter
characterizing the anisotropy of the ferromagnetic surface plasmon frequency
is evaluated to be $\eta _{S}\sim 0.01\%$.

In addition to the surfaces of ferromagnetic single layers, it has been
shown that the interface between a heavy metal and a ferromagnetic insulator
may be host to both strong SO interaction and magnetism ~\cite%
{Saitoh13PRL_SH-MR,Chien12PRL_Proximity-Pt,slZhang16PRL}. Therefore, bilayer
structures such as Pt/YIG, Au/YIG and etc. may be explored as another
platform for probing the surface ferromagnetic plasmons. Recently, spin
current generated by surface plasmon resonance was observed~\cite%
{Uchida15NatComm_plasm_Pt-YIG} in a bilayer of Pt/BiY$_{2}$Fe$_{5}$O$_{12}$
with Au nanoparticle embedded in the Pt layer, indicating the existence of a
coupling between surface plasmons and magnetic ordering in such
heterostructures.

Ferromagnetic \textit{edge} plasmons can also be hosted in various systems.
One possibility is the generation of edge plasmons in the conducting surface
of magnetic topological insulators. For instance, the quantum anomalous Hall
effect was observed in magnetically doped topological insulator (Bi,Sb)$_{2}$%
Te$_{3} $~\cite{Chang13Sci_QAHE} with quantized anomalous Hall conductivity
of $\sigma _{yx}^{QAH}\simeq \frac{e^{2}}{\hbar }$. Given this value, one
can estimate the effective ``Compton wavelength" to be $\lambda ^{2}\simeq 5$
\AA $^{2}$.~\footnote{%
The quantum Hall conductivity $\sigma _{yx}^{QAH}$ is proportional to the
Berry flux $\mathcal{F}$, i.e., $\sigma _{yx}^{QAH}=\frac{e^2}{2\pi h}%
\mathcal{F}$; given the measured value $\sigma _{yx}^{QAH}=\frac{e^2}{h}$ we
get $\mathcal{F}=2\pi$. On the other hand, the Berry flux can be evaluated
by integrating the Berry curvature $\Omega_c$ over all occupied states,
i.e., $\mathcal{F}=\int d^2\mathbf{k}\Omega_c f_{0}(\mathbf{k})$ where $%
f_{0}(\mathbf{k})$ is the equilibrium distribution function for which it is
a good approximation to replace it by a Heaviside step function $%
\Theta(k_F-k)$ when the temperature is well below the Fermi temperature.
Recalling that for our case $\Omega_c\sim \lambda^2$, we find $\mathcal{F}%
=\pi k_F^2 \lambda^2$. Therefore, we have $\lambda^2\sim \frac{2}{k_F^2}$.
Using $k_F\sim0.3$~$\mathring{A}^{-1}$ for the topological insulator, we
find $\lambda \sim 5~\mathring{A}$.}

Similar to the 3D ferromagnetic surface plasmons, we can define a quantity $%
\gamma _{E}=\left. \frac{\omega \left( q\right) -\omega \left( -q\right) }{%
\omega \left( q\right) }\right\vert _{q\rightarrow 0}$ to characterize the
chirality of the 2D ferromagnetic edge plasmons. If we use typical values of
the parameters for topological insulators~\cite{DiPietro13NatNano_plasm-TI} (%
$\omega _{q}\sim 1$ THz, $m_{e}\sim 10^{-31}$ kg), we find $\gamma _{E}\sim
0.02\%$.

Another system that may provide interesting results for the 2D ferromagnetic
edge plasmons is the 2D electron gas formed at the interface of two
dielectric perovskites, such as LaAlO$_{3}$/SrTiO$_{3}$~\cite%
{Ohtomo04NatSTO-LAO} or LaTiO$_{3}$ /SrTiO$_{3}$~\cite{Thiel06Sci_LTO/STO}.
The recent observation of ferromagnetism at LaAlO$_{3}$ /SrTiO$_{3}$
interfaces~\cite{Brinkman07NatMater_Mag-STO-LAO} is of great relevance for
the present study, providing a realization of a high-mobility magnetic 2D
electron gas.

\section{Conclusion}
\label{sec:conclusion}
A new type of surface plasmon, which does not depend on the Lorenz force but
on the spin-orbit coupling to the magnetization, has been identified. We
call it \textquotedblleft ferromagnetic surface plasmon\textquotedblright .
The frequency and the angular dependence of the ferromagnetic surface
plasmon can be controlled by varying the direction and the magnitude of the
bulk magnetization. Because the magnetization of a ferromagnetic system is a
dynamical variable with its own intrinsic oscillations (spin waves or
magnons) our results foreshadow the exciting possibility of a coupling
between spin waves and surface or edge plasmons. Such a coupling could be
exploited to control the plasmon dispersion by acting on the magnetization
via a magnetic field or a current, or, reciprocally, to induce changes in
the magnetization by pumping surface plasmons in a ferromagnetic material.
Such possibilities, if realized, could create an unexpected link between the
two apparently distant fields of plasmonics and spintronics.

\section{Acknowledgements}

We thank Alessandro Principi and Olle G. Heinonen for helpful discussions.
G.V. and S.S.-L.Z gratefully acknowledge support for this work from NSF
Grant DMR-1406568. Part of the work done by S.S.-L.Z at Argonne National
Laboratory was supported by the Department of Energy, Office of Science,
Basic Energy Sciences, Materials Sciences and Engineering Division.

\bigskip \appendix

\section{3D surface magnetoplasmon with arbitrary propagation direction}

In an earlier paper~\cite{Fetter85}, Fetter studied the dispersion of the
surface magnetoplasmons for the special case in which both the propagation
direction of the plasmon waves and the applied magnetic field are lying in
the plane of the surface and perpendicular to each other. In this appendix,
we examine the more general case where the plasmons propagate in any
arbitrary direction. To be more specific, we consider surface plasmons in a
semi-infinite metal layer occupying the space $z<0$ and the magnetic field
is applied in the $\mathbf{\hat{x}}$ direction parallel to the surface of
the metal layer.

Let us start with the following set of bulk Hydrodynamic equations, which
include the current continuity equation%
\begin{equation}
\frac{\partial \delta n}{\partial t}+\nabla \cdot \mathbf{j}=0
\end{equation}%
with $\mathbf{j}$ the number current density, the Euler equation involving
the Lorentz force term associated with the cyclotron frequency $\omega _{c}$
\begin{equation}
\frac{\partial \mathbf{j}}{\partial t}=-s^{2}\nabla \delta n+\frac{en_{0}}{m_e}%
\nabla \varphi +\omega _{c}\mathbf{\hat{x}}\times \mathbf{j} \,,
\end{equation}%
and the Poisson's equation for the electrostatic potential $\varphi $ given
by
\begin{equation}
\nabla ^{2}\varphi =4\pi e\delta n \,.  \label{Eq-App: Poisson-phi<}
\end{equation}%
Note that in the exterior of the metal layer ($z>0$) the electrons are
absent and hence the Poisson equation reduces to a Laplacian equation
\begin{equation}
\nabla ^{2}\varphi =0\,.  \label{Eq: Poisson-phi>}
\end{equation}

Assuming translational invariance in the $x$-$y$ plane, we seek solutions in
the form of plane waves of wave vector $\mathbf{q}$ but decays exponentially
along the $z$ direction towards the bulk ($z<0$), i.e., $\Psi \left( \mathbf{%
r,}z;t\right) \sim e^{\kappa z}e^{i\left( \mathbf{q\cdot r-}\omega t\right) }
$ where $\Psi $ stands for the various hydrodynamic variables under
consideration (e.g., $\delta n$, $\mathbf{j}$, $\varphi $ and etc.), $\kappa
\left( >0\right) $ is the decaying constant, and both $\mathbf{q}$ and $%
\mathbf{r}$ lie in the $x$-$y$ plane. Inserting this ansatz in the bulk
hydrodynamic equations, we find a set of linear algebraic equations for the
hydrodynamic variables, i.e.,
\begin{subequations}
\label{Eq-App:bulk-hydro}
\begin{equation}
-i\omega \delta n+\kappa j_{z}+i\mathbf{q}\cdot \mathbf{j}_{\parallel }=0
\end{equation}%
\begin{equation}
-i\omega j_{x}+iq_{x}\left( s^{2}\delta n-\frac{en_{0}}{m_{e}}\varphi
\right) =0
\end{equation}%
\begin{equation}
-i\omega j_{y}+iq_{y}\left( s^{2}\delta n-\frac{en_{0}}{m_{e}}\varphi
\right) +\omega _{c}j_{z}=0
\end{equation}%
\begin{equation}
-i\omega j_{z}+\kappa \left( s^{2}\delta n-\frac{en_{0}}{m_{e}}\varphi
\right) -\omega _{c}j_{y}=0\,,
\end{equation}%
where $\mathbf{j}_{\parallel }=\left( j_{x},j_{y}\right) $ are the in-plane
components of the current density, and the Poisson equation (\ref{Eq-App:
Poisson-phi<}) establishes a relation between $\delta n$ and $\varphi $ as
\end{subequations}
\begin{equation}
\varphi =\frac{4\pi e}{\kappa ^{2}-q^{2}}\delta n\,.  \label{Eq-App:phi-dn}
\end{equation}%
. Note that the above equations are invariant under $\mathbf{q\rightarrow -q}
$, $\omega $ $\mathbf{\rightarrow -}\omega $ and a complex conjugation, as
required by the reality of the electromagnetic waves. It is straightforward
to show that the set of equations~(\ref{Eq-App:bulk-hydro}) have nontrivial
solutions only if the following equation is satisfied%
\begin{equation}
\left( \kappa ^{2}-q^{2}\right) ^{2}-\left[ k_{\omega }^{2}-\left( \frac{%
q_{x}\omega _{c}}{\omega }\right) ^{2}\right] \left( \kappa
^{2}-q^{2}\right) -\left( \frac{q_{x}\omega _{B}\omega _{c}}{s\omega }%
\right) ^{2}=0\,,
\end{equation}%
where $\omega _{B}\equiv \sqrt{\frac{4\pi n_{0}e^{2}}{m_{e}}}$ is the 3D
bulk plasma frequency, and we have defined
\begin{equation*}
k_{\omega }^{2}\equiv \frac{\omega _{B}^{2}+\omega _{c}^{2}-\omega ^{2}}{%
s^{2}}\,.
\end{equation*}%
Note that $\kappa $ may have two positive solutions as given by
\begin{eqnarray}
\kappa _{1,2}^{2} &=&q^{2}+\frac{1}{2}\left[ k_{\omega }^{2}-\frac{%
q_{x}^{2}\omega _{c}^{2}}{\omega ^{2}}\right. \,  \notag \\
&&\left. \pm \sqrt{\left( k_{\omega }^{2}-\frac{q_{x}^{2}\omega _{c}^{2}}{%
\omega ^{2}}\right) ^{2}+\left( \frac{2q_{x}\omega _{B}\omega _{c}}{s\omega }%
\right) ^{2}}\right]   \label{Eq-App:Kappa-gen1}
\end{eqnarray}%
where $\kappa _{1}$ and $\kappa _{2}$ correspond to the solutions with the $+
$ sign and the $-$ sign respectively. Having found the decaying constants
for the hydrodynamic variables, we can now write down the general solutions
for electrostatic potential $\varphi $ and the electron density fluctuation $%
\delta n$ as follows%
\begin{equation}
\varphi =e^{i\mathbf{q\cdot r}}\times \left\{
\begin{array}{c}
\varphi _{1}e^{\kappa _{1}z}+\varphi _{2}e^{\kappa _{2}z}\text{ , \ }z<0 \\
\varphi _{0}e^{-qz}\text{ , \ \ }z>0\text{ \ \ \ \ \ \ \ \ \ \ }%
\end{array}%
\right.
\end{equation}%
and
\begin{equation}
\delta n=\frac{e^{i\mathbf{q\cdot r}}}{4\pi e}\left[ \varphi _{1}\left(
\kappa _{1}^{2}-q^{2}\right) e^{\kappa _{1}z}+\varphi _{2}\left( \kappa
_{2}^{2}-q^{2}\right) e^{\kappa _{2}z}\right] \,,
\end{equation}%
where we have invoked Eq.~(\ref{Eq-App:phi-dn}) in deriving the expression
for $\delta n$, and note that we have dropped the common time-dependent
multiplier $e^{-i\omega t}$ for ease of notation. Similarly, one can write
down the general solution for the normal component of the current density as
\begin{equation}
4\pi ej_{z}=\left( \varphi _{1}\mathcal{J}_{1}e^{\kappa _{1}z}+\varphi _{2}%
\mathcal{J}_{2}e^{\kappa _{2}z}\,\right) e^{i\mathbf{q\cdot r}},
\end{equation}%
where we have defined the quantities
\begin{equation}
\mathcal{J}_{\alpha }=i\left[ \omega _{B}^{2}-s^{2}\left( \kappa _{\alpha
}^{2}-q^{2}\right) \right] \left( \frac{\omega \kappa _{\alpha }+\omega
_{c}q_{y}}{\omega ^{2}-\omega _{c}^{2}}\right) \,
\end{equation}%
with $\alpha =1,2$.

Now by imposing the boundary conditions, i.e., the continuity of $\varphi $
and $\partial _{z}\varphi $ as well as the condition $j_{z}=0$ at the
surface $z=0$, we arrive at a general equation that determines the
dispersion of the surface magnetoplasmon%
\begin{equation}
\left( \kappa _{1}+q\right) \mathcal{J}_{2}-\left( \kappa _{2}+q\right)
\mathcal{J}_{1}=0\,.  \label{Eq-App:dipersion-gen}
\end{equation}%
Despite the complicated appearance of the general dispersion equation, its
solutions of interests in the long wavelength limit in fact can be solved
analytically. Up to $O\left( q^{1}\right) $, the two decaying constants
given by Eq.~(\ref{Eq-App:Kappa-gen1}) reduce to
\begin{equation}
\kappa _{1}^{2}\rightarrow \frac{\omega _{B}^{2}+\omega _{c}^{2}-\omega ^{2}%
}{s^{2}}  \label{Eq-App:k1-q^1}
\end{equation}%
and
\begin{equation}
\kappa _{2}^{2}\rightarrow q^{2}\left[ 1-\frac{\left( \omega _{B}\omega
_{c}\cos \phi _{q}\right) ^{2}}{\omega ^{2}\left( \omega _{B}^{2}+\omega
_{c}^{2}-\omega ^{2}\right) }\right] \,,  \label{Eq-App:k2-q^1}
\end{equation}%
where we have let $q_{x}=q\cos \phi _{q}$ and $q_{y}=q\sin \phi _{q}$. By
plugging Eqs.~(\ref{Eq-App:k1-q^1}) and (\ref{Eq-App:k2-q^1}) in Eq.~(\ref%
{Eq-App:dipersion-gen}) and making some rearrangements, we arrive at a
simpler equation for the dispersion
\begin{equation}
\left( \omega _{c}^{2}-\omega ^{2}\right) \left[ 2\omega ^{2}-2\omega
_{c}\omega \sin \phi _{q}-\left( \omega _{c}\cos \phi _{q}\right)
^{2}-\omega _{B}^{2}\right] =0\,.
\end{equation}%
Discarding the unphysical solution of $\omega =\omega _{c}$, the surface
magnetoplasmon frequency in the $q\rightarrow 0$ limit is given by
\begin{equation}
\omega _{mp}^{3D}\left( \phi _{q}\right) =\frac{\omega _{c}\sin \phi _{q}}{2}%
+\sqrt{\frac{\omega _{c}^{2}\sin ^{2}\phi _{q}}{4}+\frac{\omega
_{B}^{2}+\omega _{c}^{2}\cos ^{2}\phi _{q}}{2}}\,.
\label{Eq-App:dispersion-q0}
\end{equation}%
Interestingly, the magnetic field gives rise to a correction to the surface
magnetoplasmon frequency even when the plasmons propagate along the
direction of the magnetic field; this can be seen by setting $\phi _{q}=0$
or $\pi $ for which the dispersion relation reduces to
\begin{equation}
\omega _{mp,\parallel }^{3D}=\sqrt{\frac{\omega _{B}^{2}+\omega _{c}^{2}}{2}}%
\,.
\end{equation}%
This can be understood since it is the velocity density $\mathbf{v}$ (which
is parallel to the current density $\mathbf{j=}n_{0}\mathbf{v}$) that enters
the bulk hydrodynamic equation via the Lorentz force of the form $\mathbf{F}%
_{l}=\frac{e}{c}\mathbf{v\times H}$ with $\mathbf{H}$ the applied magnetic
field; the velocity density, however, is \textit{not} proportional to the
the wave vector $\mathbf{q}$ in the presence of the magnetic field. To see
this, it is instructive to explicitly write down the relation between $%
\mathbf{j}$ and $\mathbf{q}$ from the set of equations (\ref%
{Eq-App:bulk-hydro}) as follows
\begin{equation}
\left(
\begin{array}{c}
j_{x} \\
j_{y} \\
j_{z}%
\end{array}%
\right) =\frac{\delta n\left( s^{2}-\frac{\omega _{B}^{2}}{\kappa ^{2}-q^{2}}%
\right) }{\omega ^{2}\left( \omega _{c}^{2}-\omega ^{2}\right) }\left(
\begin{array}{ccc}
i\left( 1-\frac{\omega _{c}^{2}}{\omega ^{2}}\right)  & 0 & 0 \\
0 & i & \frac{\omega _{c}}{\omega } \\
0 & -\frac{\omega _{c}}{\omega } & i%
\end{array}%
\right) \left(
\begin{array}{c}
iq_{x} \\
iq_{y} \\
\kappa
\end{array}%
\right)
\end{equation}

For the special case considered by Fetter~\cite{Fetter85} in which the
plasmons propagate perpendicularly to the magnetic field, we can set $\phi
_{q}=\pm \frac{\pi }{2}$ in Eq.~(\ref{Eq-App:dispersion-q0}) and obtain
\begin{equation}
\omega _{mp}^{3D}\left( \pm \frac{\pi }{2}\right) =\frac{1}{2}\left[ \sqrt{%
2\omega _{B}^{2}+\omega _{c}^{2}}\pm \omega _{c}\right] \,,
\end{equation}%
where we have recovered the result for the special case derived by Fetter
(c.f., Eq.~(12) in Ref.~[\cite{Fetter85}]). By placing Eq.~(\ref%
{Eq-App:dispersion-q0}) into Eqs.~(\ref{Eq-App:k1-q^1}) and (\ref%
{Eq-App:k2-q^1}), one can verify that both $\kappa _{1}$ and $\kappa _{2}$
remain positive definite for all values of $\phi _{q}$ and $\frac{\omega _{c}%
}{\omega _{B}}$.

\bibliographystyle{apsrev4-1}
\bibliography{180410_FM-SurfacePlasmons_v3}

\end{document}